\documentclass[final,5p,times,twocolumn]{elsarticle} 

\usepackage{amssymb}
\usepackage{amsmath}
\usepackage{siunitx}
\usepackage{subcaption} 
\DeclareSIUnit[number-unit-product = {}]\nucleon{u}
\DeclareSIUnit[number-unit-product = {}]\particles{particles}
\DeclareSIUnit[number-unit-product = {}]\spill{spill}

\usepackage{color}

\usepackage{lineno,hyperref}

\journal{Nuclear Inst. and Methods in Physics Research, A}

\begin{document}
	
	\begin{frontmatter}
		
		\title{Imaging with protons at MedAustron}

		\author[add1]{F.~Ulrich-Pur}
        \author[add1]{T.~Bergauer}
		\author[add2]{A.~Burker}
		\author[add3,add4]{S.~Hatamikia}
		\author[add2]{A.~Hirtl}
		\author[add1]{C.~Irmler\corref{mycorrespondingauthor}}
		\cortext[mycorrespondingauthor]{Corresponding author}
		\ead{christian.irmler@oeaw.ac.at}
		\author[add1]{S.~Kaser}
		\author[add1]{P.~Paulitsch}
		\author[add1]{F.~Pitters}
		\author[add1,add2]{V.~Teufelhart}
        
		\address[add1]{Institute of High Energy Physics, Austrian Academy of Sciences, 1050 Vienna, Austria}
		\address[add2]{Atominstitut, TU Wien, 1020 Vienna, Austria}
		\address[add3]{Austrian Center for Medical Innovation and Technology, Wiener Neustadt, Austria}
		\address[add4]{Center for Medical Physics and Biomedical Engineering, Medical University of Vienna, Vienna, Austria}

\begin{abstract}
	Ion beam therapy has become a frequently applied form of cancer therapy over the last years. The advantage of ion beam therapy over conventional radiotherapy using photons is the strongly localized dose deposition, leading to a reduction of dose applied to surrounding healthy tissue. Currently, treatment planning for proton therapy is based on X-ray computed tomography, which entails certain sources of inaccuracy in calculation of the stopping power (SP). A more precise method to acquire the SP is to directly use high energy protons (or other ions such as carbon) and perform proton computed tomography (pCT).
	With this method, the ions are tracked prior to entering and after leaving the patient and finally their residual energy is measured at the very end.
	Therefore, an ion imaging demonstrator, comprising a tracking telescope made from double-sided silicon strip detectors and a range telescope as a residual energy detector, was set up. First measurements with this setup were performed at beam tests at MedAustron, a center for ion therapy and research in \mbox{Wiener Neustadt}, \mbox{Austria}. The facility provides three rooms for cancer treatment with proton beams as well as one which is dedicated to non-clinical research.
    \\
	This contribution describes the principle of ion imaging with proton beams in general as well as the design of the experimental setup.
	Moreover, first results from simulations and recent beam tests as well as ideas for future developments will be presented.
\end{abstract}


\end{frontmatter}


\section{Introduction}
\begin{figure}[hbt] 
	\centering 
	\includegraphics[width=1.\columnwidth,keepaspectratio]{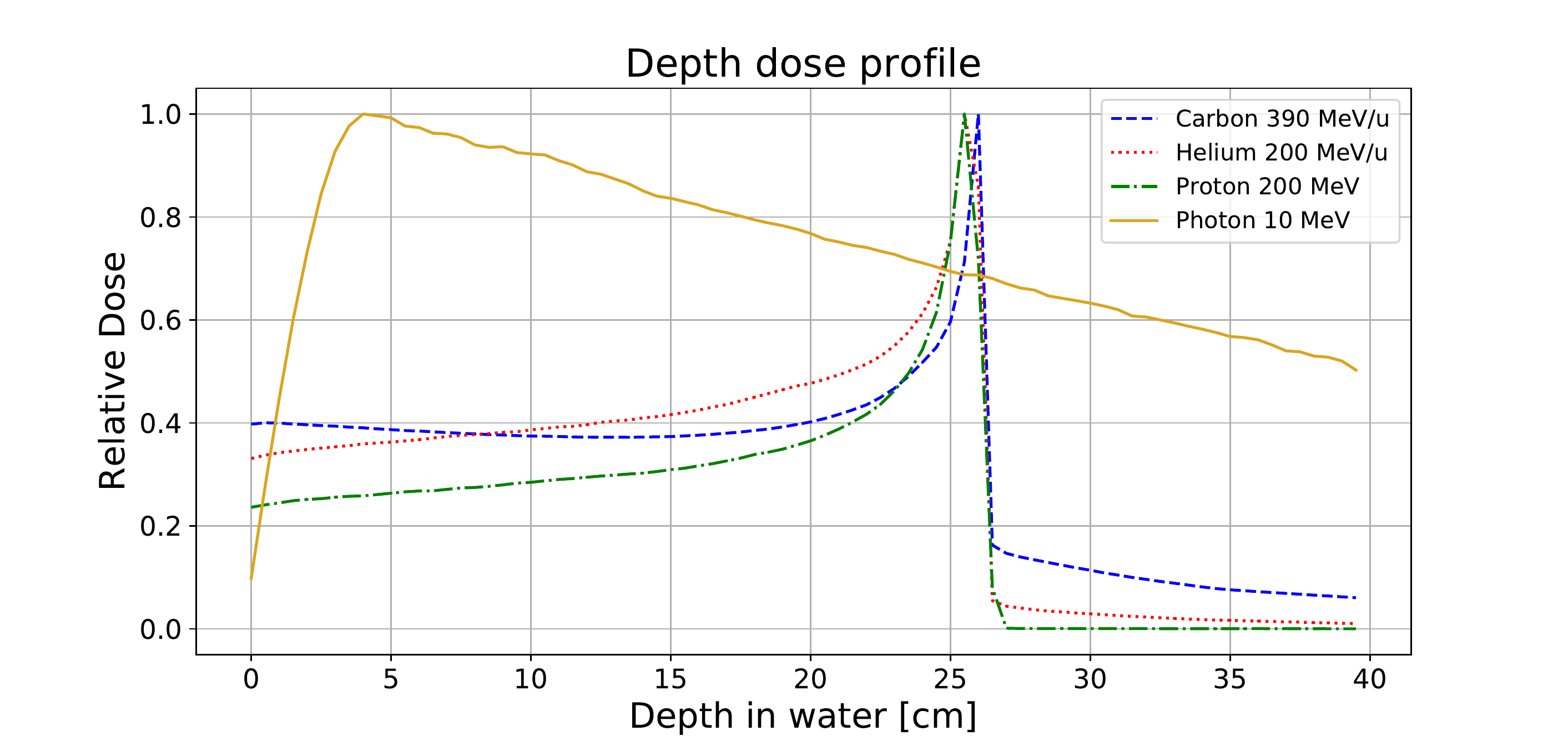}
	\caption{Monte Carlo simulation of the energy deposition of photons, protons, carbon and helium ions as a function of the penetration depth. 
	}
	\label{fig:braggpeak}
\end{figure}

Ion beam therapy is playing an increasingly important role in cancer treatment. 
The benefit of ion beams is, that ions have certain penetration lengths, depending on ion type, energy and target material, with a distinct maximum (Bragg peak) of their energy deposition at the last few millimeters of their range (Figure~\ref{fig:braggpeak}).
Compared to radiotherapy with photons, this feature of ion beams allows for strongly localized energy deposition at the target depth, while the radiation dose to the surrounding tissues is reduced.

Prior to the therapy, a treatment plan has to be established. Currently, plans are based on X-ray computed tomography (CT), characterizing the tissue in terms of Hounsfield units (HU). For ion beam therapy an extrapolation from HU to stopping power (SP)~\cite{Schneider1996} is required. 
This conversion is a major source of uncertainty leading to inaccurate determination of SP and range ~\cite{Matsufuji_1998,Schaffner_1998}.
A more suitable approach is to use a high energy proton beam, which traverses the patient, to directly determine the 
SP distribution by performing a proton computed tomography (pCT)~\cite{Schulte2003}. 

\section{Experimental Setup }
\begin{figure}[hbt] 
	\centering 
	\includegraphics[width=\columnwidth,keepaspectratio]{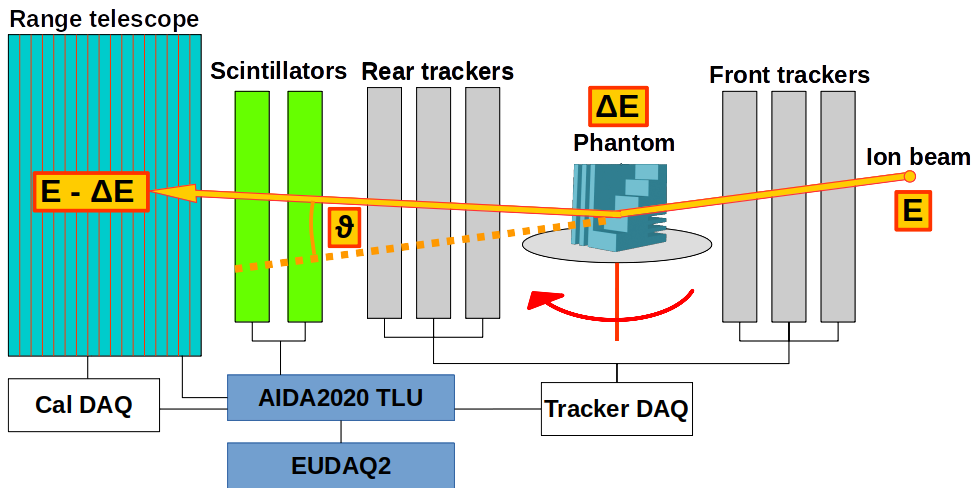}
	\caption{Sketch of the used experimental imaging setup.}
	\label{fig:pCT_setup}
\end{figure}

A pCT setup as depicted in Figure~\ref{fig:pCT_setup} consists of two particle tracker elements in front of and behind the patient to determine the tracks of the passing protons as well as a residual energy detector. 
The SP is determined from the energy deposition of the particle along its path through the patient, given by the difference between initial and residual energies.
The two trackers provide position and direction of the proton when it enters and leaves the patient and thus allow to reconstruct the most likely path~\cite{MLP} of the proton through the patient. Performing such a radiography for several incident angles and combining the data of tracker and residual energy detector is then used to determine a 3D distribution of the relative stopping power. 

The long-term goal of this project is to build a pCT system for clinical application. In a first step a demonstrator of an imaging system with three front and three rear tracker planes and a range telescope was set up and subsequently tested in several beam tests with protons of various energies at MedAustron, a cancer treatment facility located in Wiener Neustadt, Austria. MedAustron provides proton beams from \SIrange{62.4}{252.7}{\mega\electronvolt} as well as carbon ion beams from \SIrange[per-mode=symbol]{120}{402.8}{\mega\electronvolt\per\nucleon} for ion beam therapy. The facility features three rooms for treatment and one dedicated to non-clinical research room, where a proton beam with up to \SI{800}{\mega\electronvolt} can be provided.


\subsection{Tracker}
Six modules equipped with double-sided silicon strip detectors (DSSDs) were used for the tracker.  
The sensors are made from n-substrate silicon with a thickness of \SI{300}{\micro\meter} and an active area of \SI[product-units=power]{25 x 50}{\milli\meter}. They feature 512 AC coupled strips on each side, which are arranged orthogonally at a pitch of \SI{50}{\micro\meter} on the p-side ($y$ coordinate) and \SI{100}{\micro\meter} on the n-side ($x$ coordinate), respectively.
In contrast to single-sided silicon strip detectors as used for the PhaseII pCT scanner~\cite{bashkirov2016phaseII} and the PRaVDA pCT system~\cite{esposito2018PRaVDA}, DSSDs provide two-dimensional track points with a single sensor and thus allow to reduce multiple Coulomb scattering within the tracker planes. 
The DSSDs used to build the tracker modules have been originally designed for the Belle II Silicon Vertex Detector \cite{valentan2013diss} and were chosen due to their availability. For a clinical application, the size is not sufficient and the spatial resolution is over designed \cite{Johnson_2017}. In a future iteration, it is therefore planned to increase the size of the sensors while keeping the number of readout channels constant.
\begin{figure}[hbt] 
	\centering 
	\includegraphics[width=.9\columnwidth,keepaspectratio]{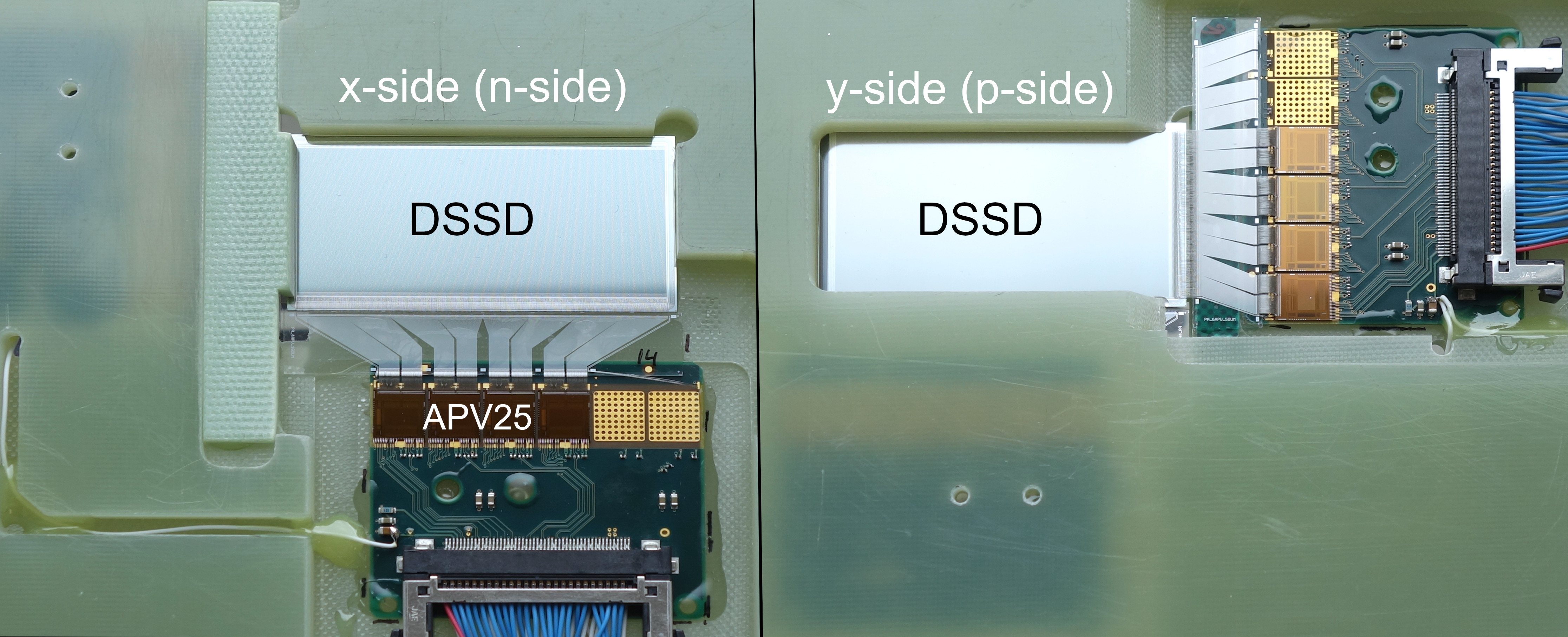}
	\caption{View of n-side and p-side of a tracker module showing the DSSD and the front-end electronics with the APV25 chips.}
	\label{fig:tracker_module}
\end{figure}

\begin{figure}[hbt] 
	\centering 
	\includegraphics[width=\columnwidth,keepaspectratio]{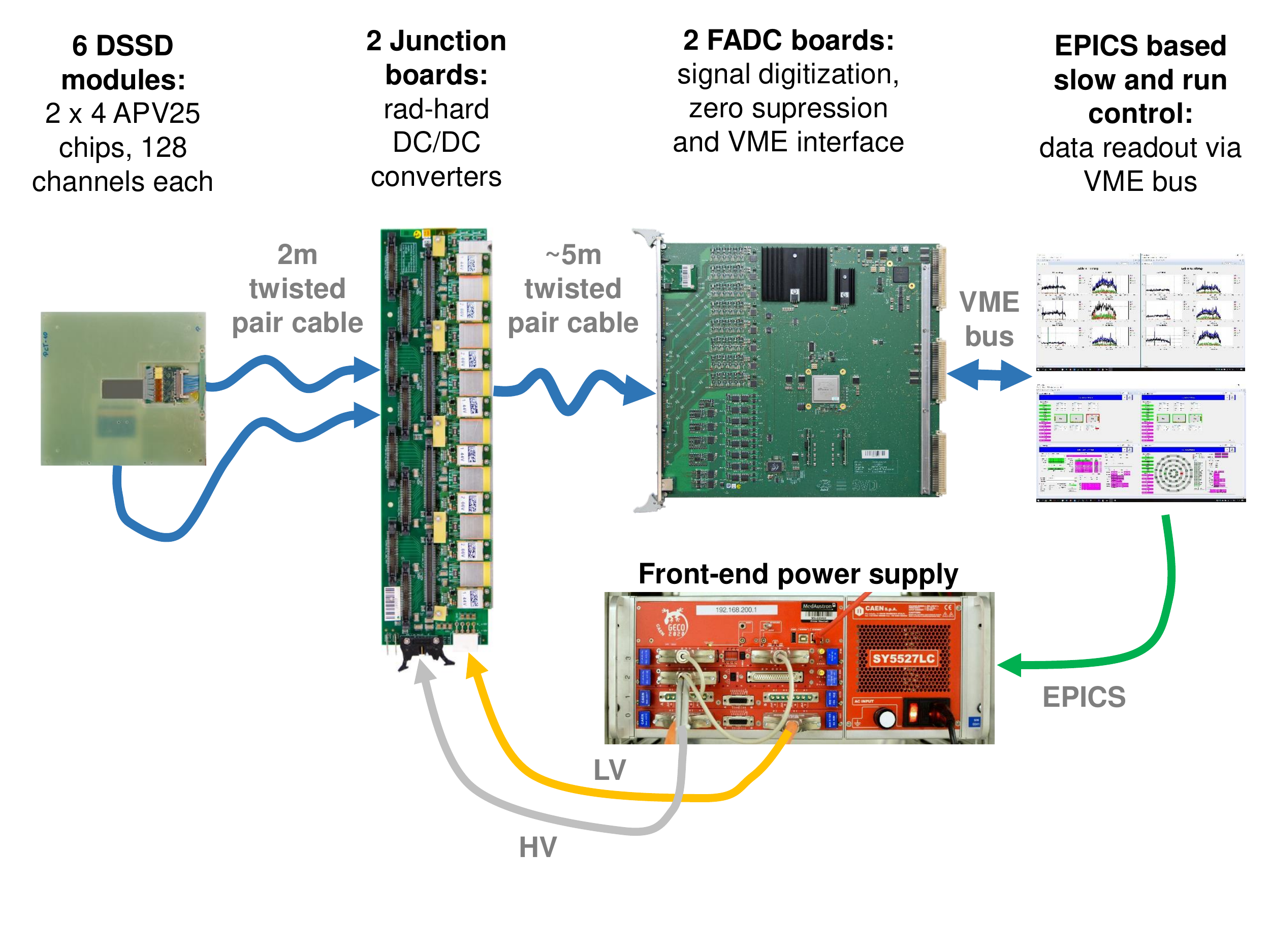}
	\caption{Readout chain of the DSSD tracker.}
	\label{fig:tracker_readout}
\end{figure}

The sensors are glued onto support frames made from glass-reinforced epoxy laminate (FR-4) and the strips on each side are connected to four APV25~\cite{APV25} chips located on hybrid boards next to the sensor (Figure~\ref{fig:tracker_module}). The chips are read out by a VME based flash ADC (FADC) system, originally developed and used for the Belle~II~Silicon~Vertex~Detector~\cite{BelleIISVDReadout}. The full read out chain is shown in Figure~\ref{fig:tracker_readout}. The DSSD modules are connected to junction boards, providing the required front-end voltages via DC/DC converters as well as the bias voltage of the sensors. The analog signals are then transmitted via \SI{5}{\meter} long twisted pair cables to the FADC boards, where they are digitized and zero suppressed. Finally, the data are read out by the EPICS based run and slow control software~\cite{Irmler_2019}, which in addition controls the CAEN power supply. The data transfer from the FADC boards to the data acquisition PC is currently implemented via a VME bus interface, which allows a data acquisition (DAQ) rate of up to \SI{500}{\hertz}. 

\subsection{Range telescope}
A proton range telescope, formerly developed by the TERA foundation \cite{Bucciantonio} was used as a residual energy detector.
It consists of 42 plastic scintillator slices with an active area of \SI[product-units=power]{300 x 300}{\milli\meter} and thickness of \SI{3}{mm}. The plastic scintillators are coupled to \SI{1}{\square\milli\meter} silicon photomultipliers (SiPM), which are attached to a custom DAQ board, where the signals are digitized and subsequently read out by a {LabVIEW} based software \cite{Bucciantonio}. 
This range telescope allows to measure protons with energies up to $\approx$~\SI{140}{\mega\electronvolt} at a data acquisition rate of $\leq$~\SI{1}{\mega\hertz}.


\subsection{Imaging setup}
In order to synchronize the data obtained by the tracker and the range telescope, the AIDA2020 trigger and logic unit (TLU) \cite{Baesso_2019}, implemented in the EUDAQ2 framework \cite{Liu_2019}, was used. The coincident signal of two \SI[product-units=power]{50x50x10}{\milli\meter} plastic scintillators, located between the rear tracker and the range telescope and connected to the TLU was used as a trigger.

\begin{figure}[hbt] 
	\centering 
	\includegraphics[width=0.5\columnwidth,keepaspectratio]{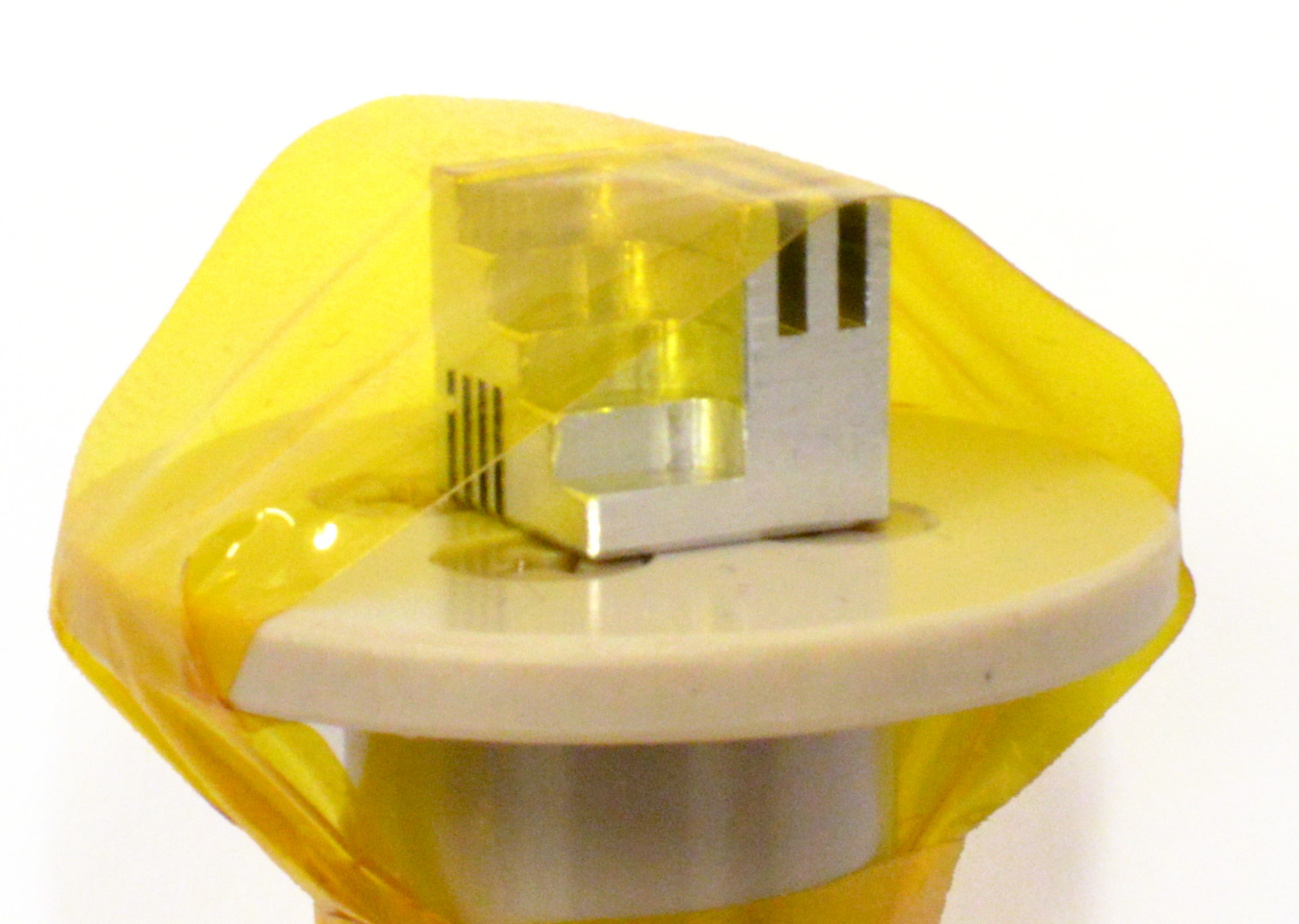}
	\caption{Stair shaped aluminum phantom taped to a rotatable mount table.}
	\label{fig:phantom_taped}
\end{figure}

\begin{figure}[hbt] 
	\centering 
	\includegraphics[width=\columnwidth,keepaspectratio]{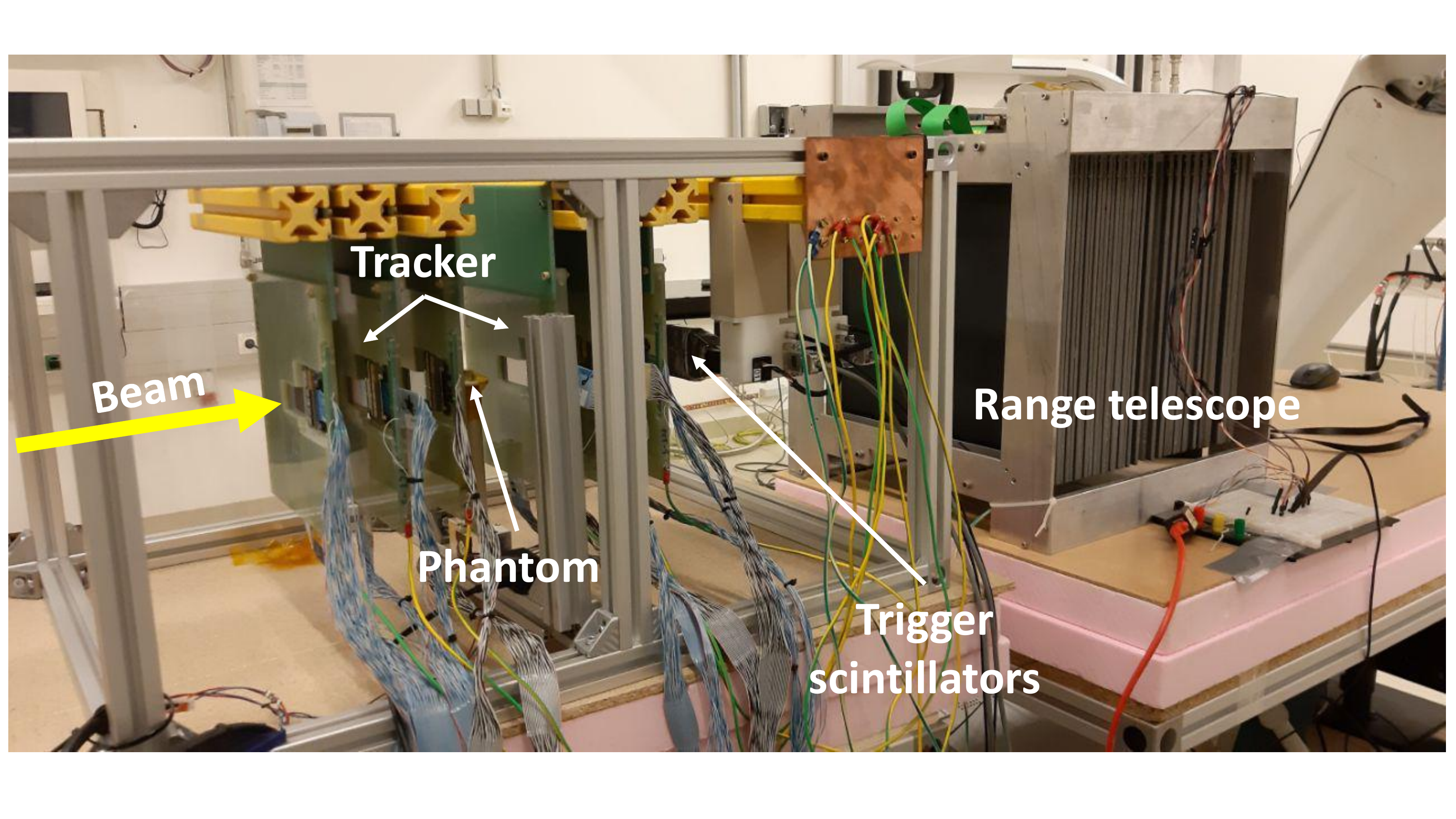}
	\caption{Experimental setup used at MedAustron.}
	\label{fig:Nov19_setup}
\end{figure}

A schematic overview of the imaging setup is depicted in Figure~\ref{fig:pCT_setup}. The object to be imaged (phantom) is placed upon a rotating table between the front and rear trackers and irradiated from various angles.
The phantom itself (Figure~\ref{fig:phantom_taped}) is a \SI{1}{\cubic\centi\meter} aluminum cube with \SI{2}{\milli\meter} steps and cutouts with \SI{0.5}{\milli\meter} and \SI{1}{\milli\meter} width.
An image of the experimental setup is shown in Figure~\ref{fig:Nov19_setup}.


\section{Imaging Methods}
\label{sec:reco}
\subsection{pCT reconstruction workflow based on simulated data}
\begin{figure}[hbt] 
	\centering 
	 \includegraphics[width=\linewidth]{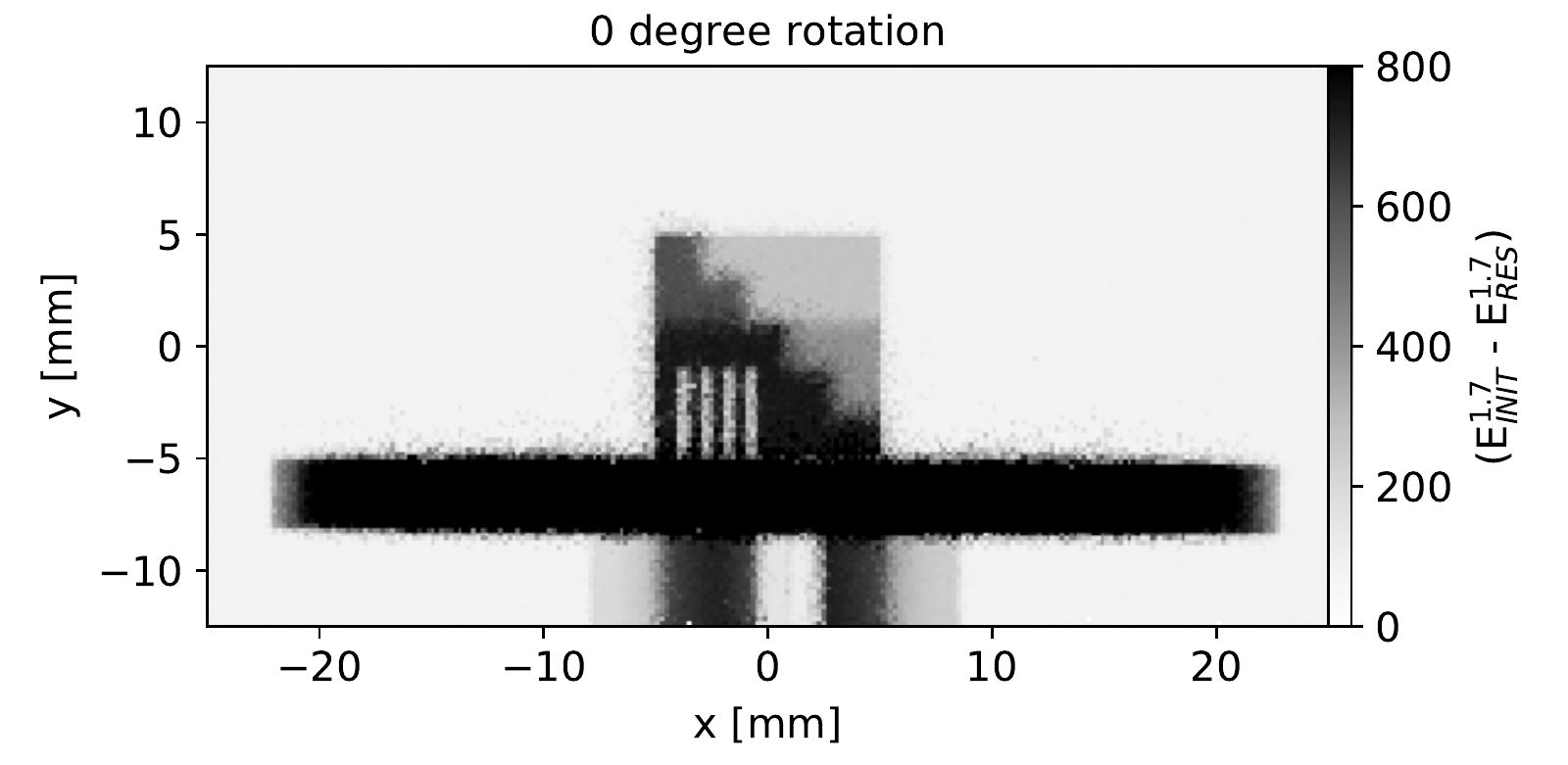}
    	 \includegraphics[width=\linewidth]{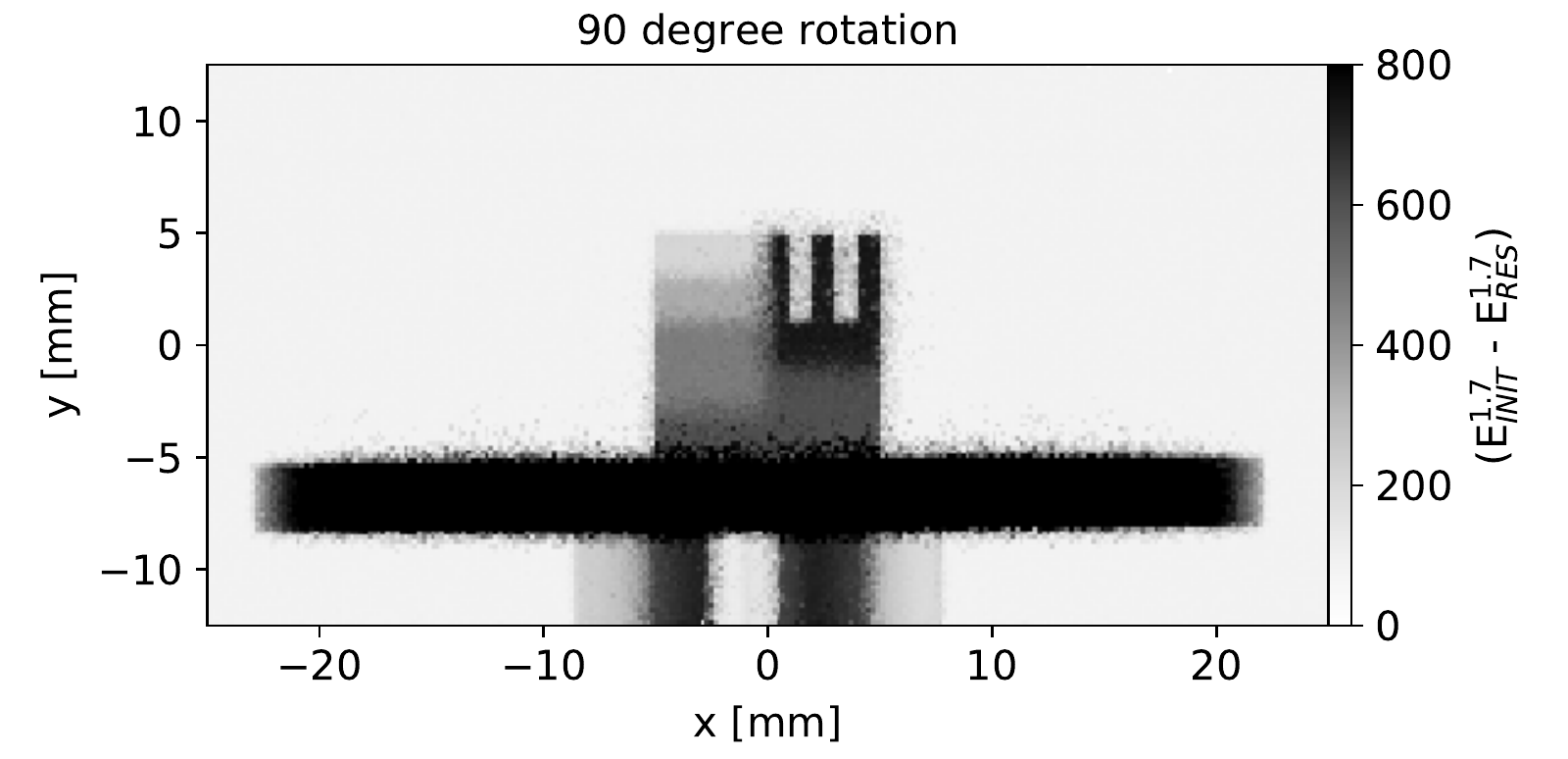}
	\caption{Simulated pCT projections at 0$^{\circ}$ and 90$^{\circ}$. Using a straight line approach calculated as the mean of proton hit positions on the tracking planes directly prior to and after the phantom, the residual proton energy (taken from the detector after the phantom) was assigned to a pixel on a plane perpendicular to the beam direction.}
	\label{fig:projections_E}
\end{figure}

In pCT, 3D information on the spatial structure and stopping power within a phantom can be obtained by irradiation from several angles. For this purpose, 2D forward projections are recorded by assigning the energy loss of each proton, which is obtained from residual energy measurements, to a certain position (pixel) on a plane perpendicular to the beam direction by using the tracker measurements. 
Since so far, no full dataset from the experimental setup is available, a Geant4  \cite{Agostinelli2002hh} Monte Carlo model was used to simulate this process in order test the following proposed reconstruction workflow. 180 projections of the aluminum cube shown in Figure~\ref{fig:phantom_taped} using \num{5e5} protons per projection in steps of 1$^{\circ}$ were generated in the simulation. Ideal spatial and energy resolutions of the detectors prior to and after the object were assumed and the initial beam energy was set to 100.4 MeV. Resulting projections simulated at 0$^{\circ}$ and 90$^{\circ}$ can be seen in Figure~\ref{fig:projections_E}.

A lot of investigation in the field of pCT image reconstruction was already performed in \cite{rit2013filtered,penfold2010block,hansen2016fast}. For this project the MATLAB/CUDA based framework TIGRE (Tomographic Iterative GPU-based REconstruction toolbox) was chosen as an initial fast and easy-to-apply solution to reconstruct the 3D image.
This framework already offers a set of reconstruction algorithms for CT from four main algorithms families: filtered back projection, simultaneous algebraic reconstruction technique (SART) type, the Krylov subspace method and the total variation regularization \cite{biguri2016tigre}. In order to use this code without modification, straight-line proton paths inside the phantom have been assumed in the reconstruction. This first-order approximation ignores the multiple Coulomb scattering (MCS) of the protons inside the phantom. The most accurate path estimates for pCT have been shown to be cubic spline and most likely path \cite{wang2011bragg}.

The Bragg-Kleeman rule \cite{bragg1905xxxix} 

\begin{equation}
-\frac{dE}{dx} = -\frac{E^{1-p}}{p \alpha},
\label{eq:bragg_kleeman}
\end{equation}

was used to approximate the stopping power within the phantom since it contains the energy-independent material parameter $\alpha$ which can be extracted from the reconstructed image. $E$ is the proton energy and $p$ is set to \num{1.7} for protons at the considered energy in aluminum \cite{zhang2010water}. In order to obtain $\alpha$, the value for $p$ is inserted in Equation (\ref{eq:bragg_kleeman}) which then transforms to

\begin{equation}
 	1.7 \cdot E^{0.7} \text{d}E = \frac{1}{\alpha(x,y,z)} \text{d}z,
	\label{eq:bragg_kleeman2}
\end{equation}

where $\text{d}z$ is an infinitesimal path element along the assumed proton path in $z$-direction. Equation (\ref{eq:bragg_kleeman2}) is integrated over proton energy and path,

\begin{equation}
 	1.7 \cdot \int_{E_{\text{init}}}^{E_{\text{res}}} E^{0.7} \text{d}E = \int_{z_{\text{in}}}^{z_{\text{out}}} \frac{1}{\alpha(x,y,z)} \text{d}z,
	\label{eq:reco_forward}
\end{equation}

where $E_{\text{init}}$ is the initial proton beam energy, $E_{\text{res}}$ is the residual proton energy and $z_{\text{in}}$ and $z_{\text{out}}$ are the proton entry and exit position to the phantom.  Solving the left side of Equation (\ref{eq:reco_forward}) and approximating the path integral by a sum, the forward projection can finally be defined as the left side of

\begin{equation}
 	E_{\text{init}}^{1.7}-E_{\text{res}}^{1.7} \approx -\sum \frac{1}{\alpha (x,y,z)} {\Delta}z,
\end{equation}

to obtain the reciprocal of the unknown parameter $\alpha$. This step is analogous to the projection definition in X-ray CT, where the unknown function is the absorption coefficient of a material.

\subsection{Multiple scattering radiography}
In order to have an image reconstruction for material estimation without depending on residual energy measurement, a second reconstruction workflow was applied. 
Clusters from hits on the tracking planes were used to create track based multiple scattering radiographies \cite{SchuetzeJansen}, using beam test data from a proton beam at MedAustron with an initial kinetic energy of \SI{100.4}{\mega\electronvolt} \cite{Burker_2019}. 
Two projections of the phantom were acquired with approximately \num{3.5e5} tracks per projection, with a large enough spot size to completely cover the phantom.

The clusters were grouped by the front and rear trackers to create two linear track segments which meet at a point of closest approach. 
A plane normal to the beam direction was partitioned into \SI[product-units=power]{500 x 500}{\micro\meter} pixels in the \(x\)- and \(y\)-direction and located at the z-position of the phantom. 
Each track was associated with a pixel in this plane. 
Thus, each pixel was linked to a distribution of kink angles. 
For each individual particle the kink angle is defined as the change in angle between the front and rear track segments, projected onto the \(x\)- and \(y\)-directions.
Both of these projected angles were then combined to obtain

\begin{equation}
    \theta = \sqrt{
        (\theta_\text{rear}^{x} - \theta_\text{front}^{x})^2 +
        (\theta_\text{rear}^{y} - \theta_\text{front}^{y})^2}.
\end{equation}

For each pixel, the median of this distribution of combined angles was used as gray scale value in the forward projection image. 

\section{Results}

\label{sec:results}

\subsection{Calibration of the range telescope}
\begin{figure}[hbt]
	\flushleft
	\includegraphics[width=0.45\textwidth]{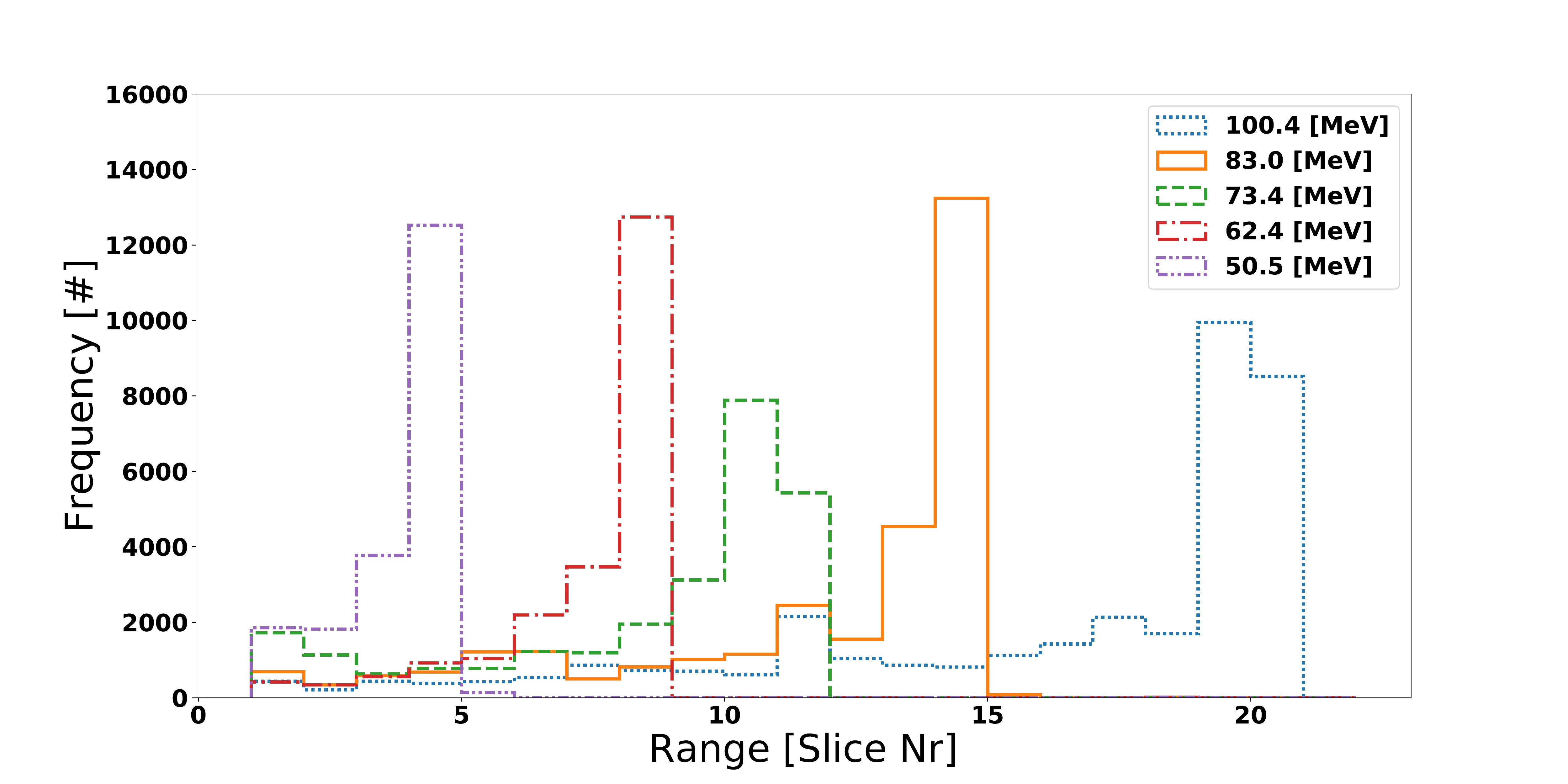}
	\caption{Range measurement for different energies. A high efficiency loss due to a missing signal at the Bragg peak position is shown.}
  \label{fig:measrange}
	\end{figure}

\begin{figure}[hbt]
  
  \centering
  \includegraphics[width=0.99\columnwidth,keepaspectratio]{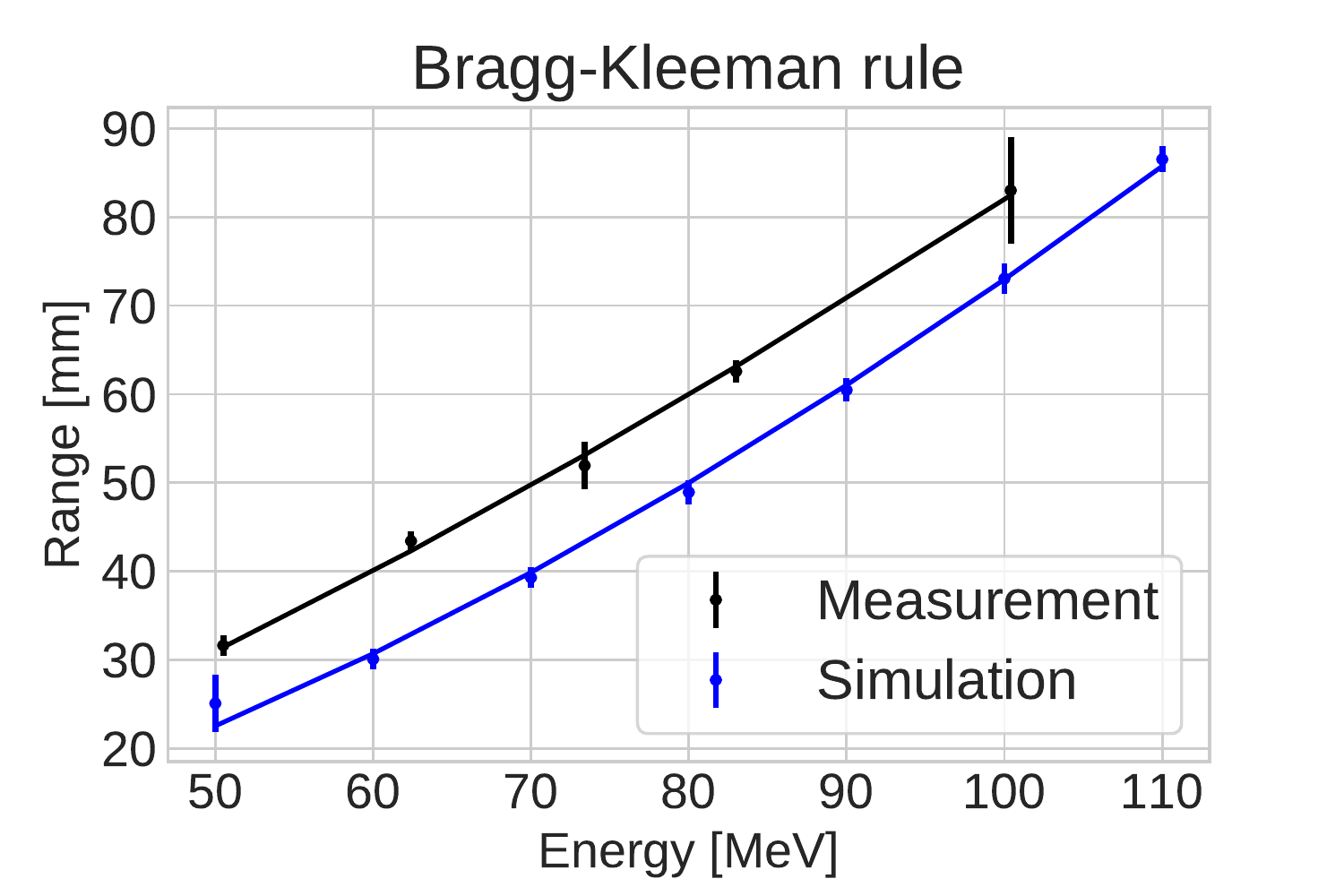}
  \caption{Geant4 simulation of the range calibration curve of the TERA range telescope compared to the measured range for different energies, using a $\SI{1.5}{MeV}$ threshold.}
  \label{fig:braggklfit}

\end{figure}

After calibrating the gain of the SiPMs with $\SI{800}{MeV}$ protons at MedAustron, the range for different proton energies was measured, using $\SI{5E5}{}$ events per energy (Figure~\ref{fig:measrange}). Due to instabilities and hardware failures of the SiPM voltage supply, only 20 slices could be calibrated.\\
To obtain the range, the acquired ADC values per scintillator slab were converted to deposited energy for each slice. The particle range was then defined as the position of the last slice over a certain threshold. Figure \ref{fig:measrange} shows the obtained ranges for energies between $\SI{50.5}{}$ and $\SI{100.4}{MeV}$. A \SI{3}{\centi\meter} thick polymethyl methacrylate slab was placed in front of the range telescope to reduce the energy of \SI{83}{} and \SI{100.4}{\mega\electronvolt} proton beams to \SI{50.5}{} and \SI{73.4}{\mega\electronvolt}, respectively. \\
In Figure \ref{fig:braggklfit} the measured ranges for various proton energies, with an energy threshold of $\SI{1.5}{MeV}$, were compared to a Geant4  simulation using the range definition obtained from the Bragg-Kleeman rule \cite{bragg1905xxxix} 
\begin{equation}
R=\alpha {E}^{p}.
\label{eq:braggkleeman}
\end{equation}
For this measurement, a systematic difference of $\approx \SI{1}{cm}$ in range as well as an efficiency loss of $\approx 95\%$ was observed. 
The cause of these issues is still unknown and under active investigation.
Because of this low efficiency and the instabilities of the SiPM voltage supply, a full pCT reconstruction was only applied to simulated data of this pCT setup.

\subsection{pCT reconstruction workflow with simulated data}
\begin{figure}[hbt] 
	\centering 
	\includegraphics[width=\linewidth]{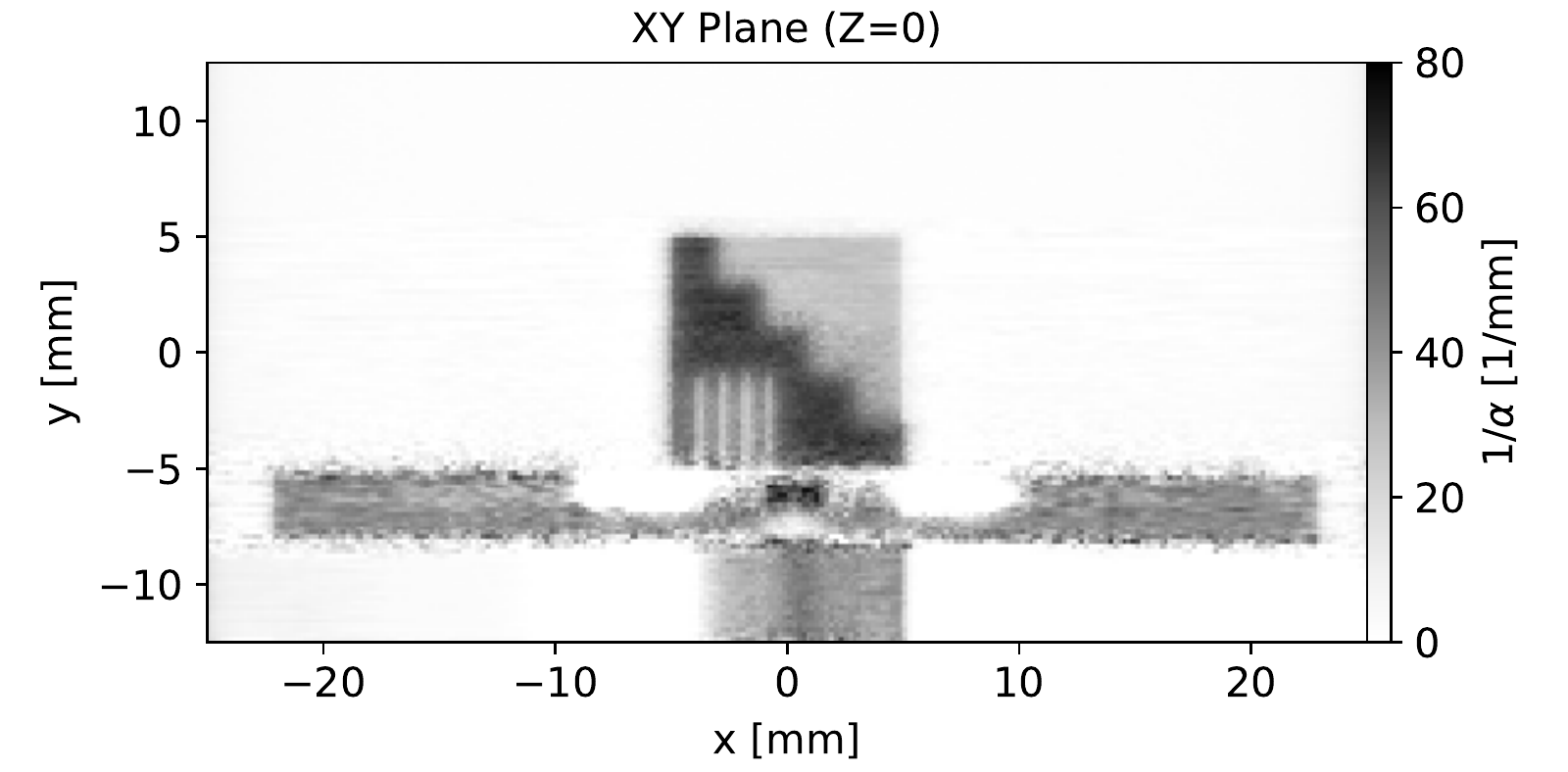}
        \includegraphics[width=\linewidth]{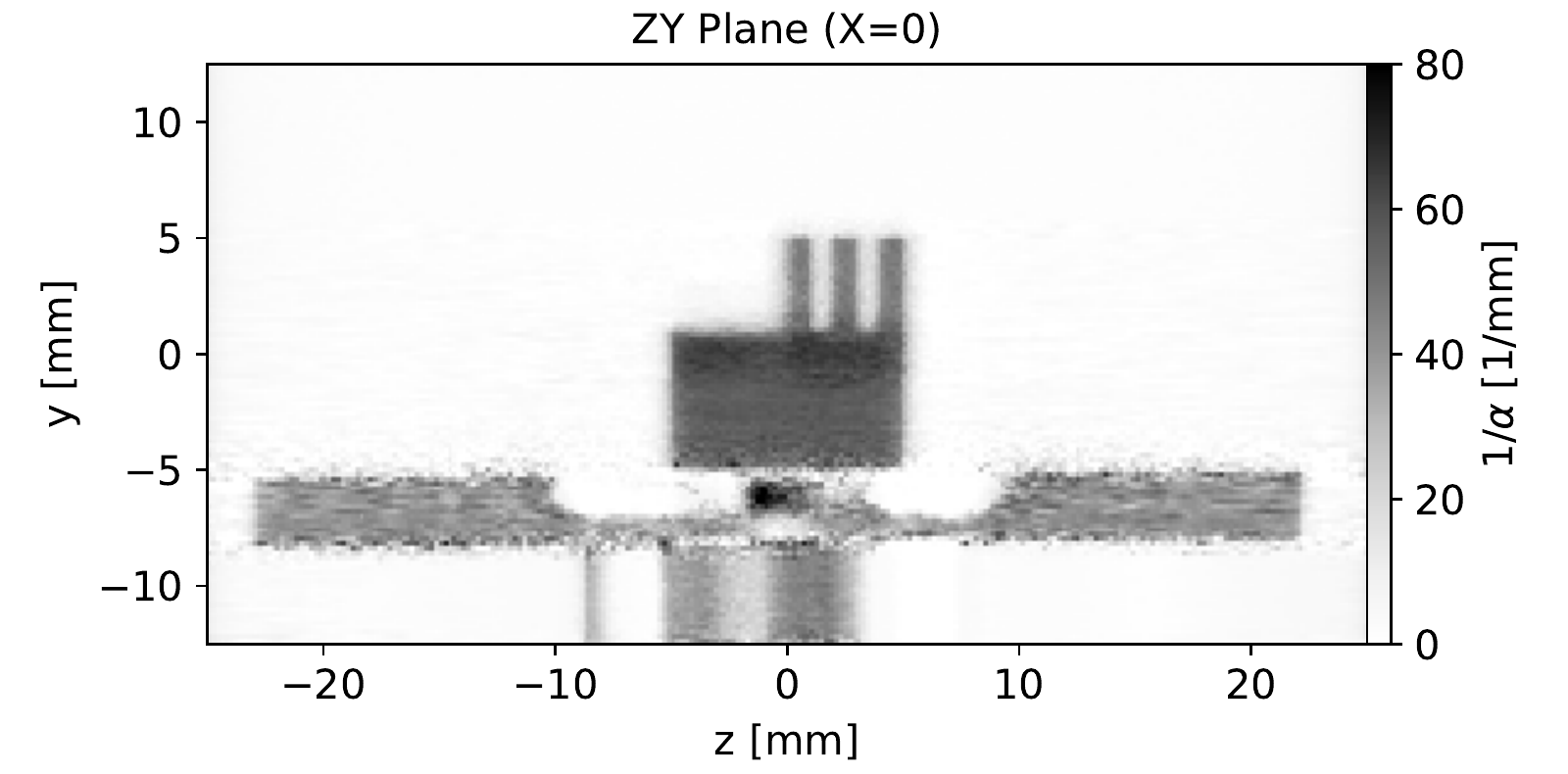}
	\caption{Preliminary reconstruction results (cuts through center of object). The gray values are given by the reciprocal of $\alpha$, which can be used to determine the SP.}
	\label{fig:reco_results}
\end{figure}

The full pCT image reconstruction chain is still work in progress. However, with the presented frameworks and assumptions made in Section~\ref{sec:reco}, preliminary results can be obtained using simulated data. These results cannot be used in order to evaluate the real experimental setup and its accuracy for image reconstruction. Nevertheless, the simulated data were used for a first test of the reconstruction workflow itself (and its respective assumptions and approximations). The iterative algorithm OS-SART \cite{censor2002block} (a member of the iterative SART-type algorithm family) of the TIGRE toolkit performed best regarding reconstruction time (less than \SI{10}{s} for 5 iterations on a Nvidia GeForce GTX 1080 Ti) and stopping power accuracy. A volume of \(15 \times 15 \times 15\) voxels with a voxel size of \( 0.2 \times 0.2 \times \SI{0.2}{\cubic\milli\meter}\) has been used to determine the SP in a region of interest (ROI) within the reconstructed phantom. Compared to a SP literature value of aluminum \cite{berger2017stopping} of \SI{15.28}{\mega\eV\per\cm} at \SI{100.4}{\mega\eV}, the observed average value of \SI{15.13}{\mega\eV\per\cm} results in a relative error of approximately \SI{1}{\percent}. In Figure~\ref{fig:reco_results}, two sectional views of the reconstructed image of the aluminum phantom can be seen. 

\subsection{Multiple scattering radiography}

\begin{figure}[hbt] 
    \centering 
    \includegraphics[width=\linewidth]{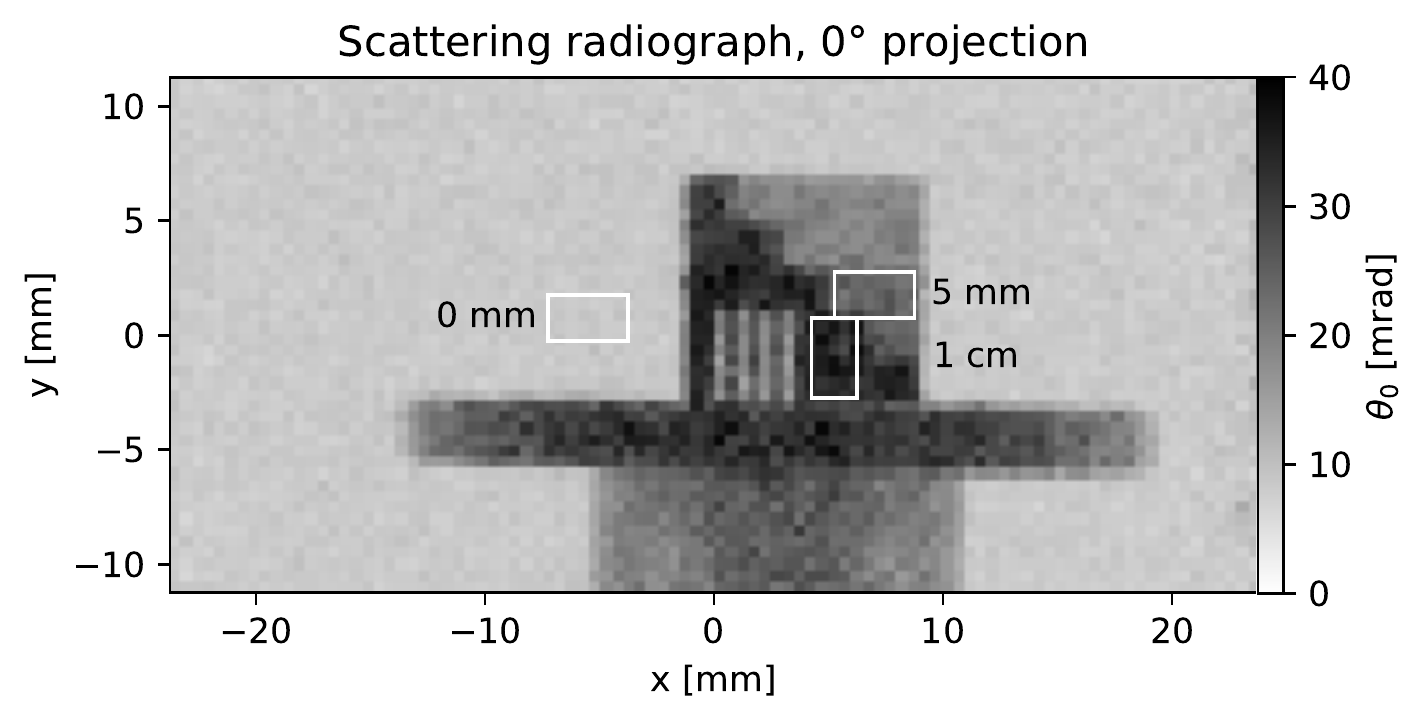}
    \includegraphics[width=\linewidth]{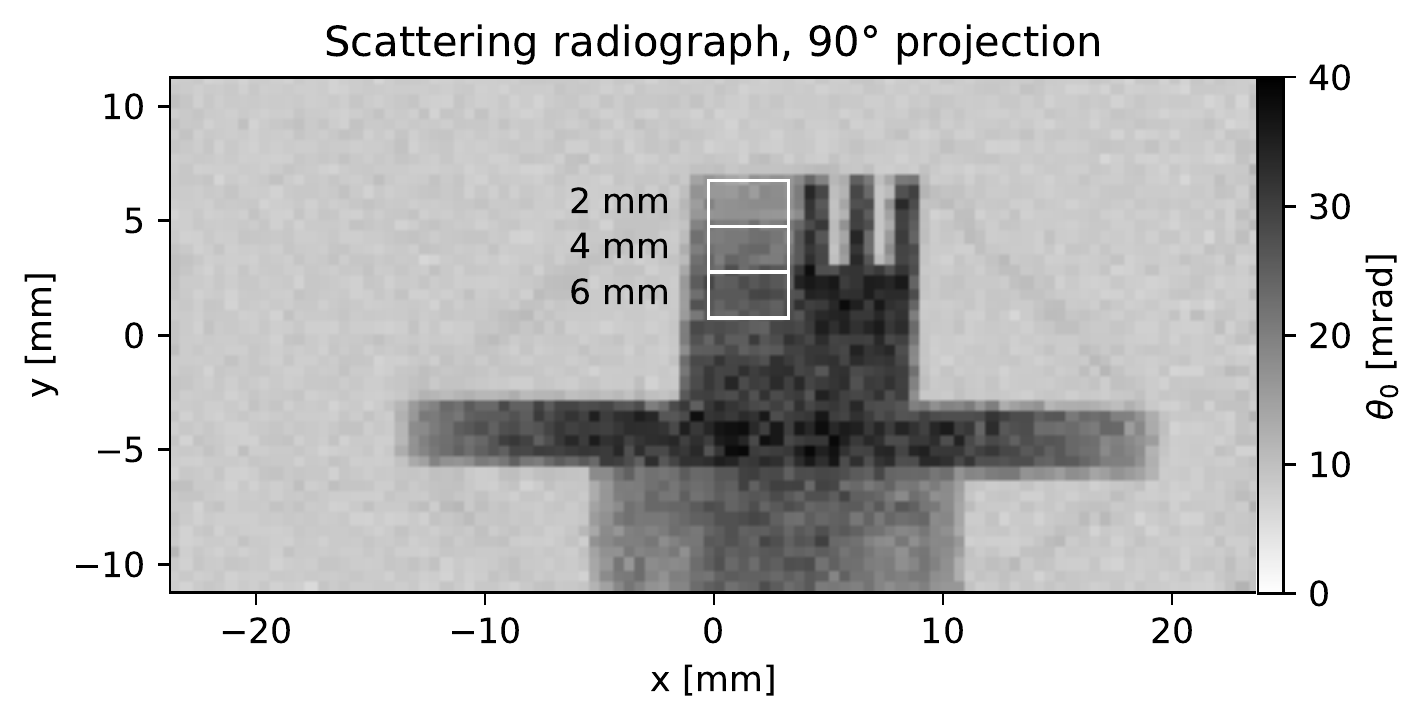}

    \caption{\label{fig:MSC_projections} Measured scattering radiographs at two phantom rotations: \ang{0} and \ang{90}.
    Six regions of interest are marked as rectangles and annotated with the known thickness of aluminum in the \(z\)-direction. }
\end{figure}

By applying the track-based multiple scattering method  \cite{SchuetzeJansen} on the tracker data from beam tests at MedAustron, images of the position resolved widening of a proton beam due to multiple Coulomb scattering were obtained.
Figure~\ref{fig:MSC_projections} illustrates this for two different rotations of the stair profile phantom.
Six regions with a known material thickness were selected to investigate the accuracy of the scattering estimates.
These regions are indicated in Figure \ref{fig:MSC_projections} as white rectangles, annotated with the corresponding thickness of aluminum in the \(z\)-direction.
A thickness of \SI{0}{\milli\meter} corresponds to no phantom and defines the background due to scattering in the silicon sensors.
Kink angles within these regions were compared to the expectation given by the Highland approximation according to  \cite{LYNCH1991}

\begin{equation}
\theta_0 = 
\frac{\SI{13.6}{\mega\electronvolt}}{\beta c p} z \sqrt{\frac{x}{X_0}}
\left[1 + 0.038 \ln\left(\frac{x}{X_0}\right)\right],
\end{equation}

where \(\beta c\), \(p\) and \(z\) are the proton's velocity, momentum and charge number, respectively, \(x\) is the thickness of the phantom and \(X_0 = \SI{88.97}{\milli\meter}\) the radiation length of aluminum \cite{PDGAtomicNuclearProperties}. 

Proton energy loss in aluminum was numerically evaluated.
The total depth of \SI{1}{\centi\meter} was subdivided into many equally thin slices in which the energy loss is almost constant.
Given an initial energy \(E_0\), this enabled the determination of the local energy at different depths

\begin{equation}
    E_i = E_{i - 1} - S(E_{i - 1}) \times dx,
\end{equation}
 
where \(E_i\) is the energy at the \(i\)-th slice, \(dx\) the slice thickness and \(S(E_{i - 1})\) is the stopping power, which was calculated with the Bethe formula \cite{Bethe1930} and the energy of the previous slice.
The geometric mean \(\sqrt{E_0 E_I}\) of initial and final energy \(E_I\) was used to calculate momentum \(\beta c\) and velocity \(p\) for the expected scattering angle according to the Highland formula. 

For the background, the Highland model was evaluated with the energy loss in \(x = 6 \times \SI{300}{\micro\meter}\) of silicon (radiation length \(X_0 = \SI{93.7}{\milli\meter}\) \cite{PDGAtomicNuclearProperties}). 
The mean background was subtracted from the mean values of the other regions, while the standard deviation of the background was added to those of the others.
This allowed us to obtain the difference in scattering due to the aluminum phantom only (Table \ref{tab:expectedVsObservedAngles}).
Available data are in good agreement with the expected amount of scattering, with systematic differences smaller than the measurement uncertainty. 
Standard deviations were found between \SIrange{2}{3}{\milli\radian} and could be reduced by recording a larger amount of proton histories per projection.

\begin{table}
    \centering
    \caption{\label{tab:expectedVsObservedAngles} Expected and observed scatter distribution widths for each material thickness.
    Mean measurement angles were reduced by the mean background, while the standard deviations were added together. }

    \begin{tabular}{
            c
            S[table-format=2.2]
            S[table-number-alignment=center,
                table-figures-integer=2,
                table-figures-decimal=2,
                table-figures-uncertainty=1,
                separate-uncertainty = true]
        }
        Thickness \([\si{\milli\meter}]\) & \text{Expectation} [\si{\milli\radian}] & \text{Observation} [\si{\milli\radian}] \\
        \hline
        background & 8.48 & 8.54 \pm 0.57 \\
        2 & 9.27 & 9.42 \pm 2.26 \\
        4 & 13.73 & 14.05 \pm 2.60 \\
        5 & 15.62 & 16.26 \pm 2.92 \\
        6 & 17.40 & 17.54 \pm 2.07 \\
        10 & 23.81 & 25.21 \pm 2.92 \\
    \end{tabular}
\end{table}

\section {Summary and Outlook}

A demonstrator of an ion imaging system, using six tracker
planes made of double-sided silicon strip detectors and 42 plastic scintillators 
to be used as a range telescope is presented. Efforts are being made to synchronize these
otherwise independent systems with a trigger logic unit in order to correlate
proton path and energy loss data. 
Since the tracker modules are read out via VME bus interface, the DAQ rate is currently limited to \SI{500}{\hertz}. In order to achieve higher DAQ rates a data transfer based on user datagram protocol (UDP) via Gigabit Ethernet interface is currently being implemented. 

The constructed demonstrator has been used at beam tests in July and November
2019. Due to hardware instabilities and high efficiency loss of the range telescope, only the tracking data are currently available in a useful quality.
Hardware upgrades for stabilizing and monitoring of the SiPM voltages of the range telescope, as well as other calorimeter technologies are currently under investigation.

The obtained tracking data were used to create track-based multiple scattering radiographies of
an aluminum stair phantom with cutouts. Measured beam widening due to multiple
Coulomb scattering was compared to estimates with the Highland formula and a
simple energy loss computation. Kink angles were overestimated by a few percent,
with an increasing discrepancy for larger phantom thicknesses. These errors were
likely caused by the simplistic treatment of energy loss as a geometric mean 
instead of an integral.

A replica of the physical setup was modelled in the Geant4 simulation framework to generate auxiliary data.
These are being used to explore available toolkits for data analysis, with respect to image reconstruction, and to prepare a common reconstruction workflow for relative stopping power and scattering power imaging with nonlinear path models taken into account.

\section*{Acknowledgements} 
The authors would like to thank A.~Bauer, W.~Brandner, S.~Schultschik, B.~Seiler, R.~Stark, H.~Steininger, R.~Thalmeier and H.~Yin for their contributions to the construction of the ion imaging demonstrator. 
This project received funding from the Austrian Research Promotion Agency (FFG), grant numbers 875854 and 869878.

\bibliography{PaperHiroshima}

\end{document}